\documentclass[11pt,letterpaper]{article}

\usepackage{xr}
\usepackage[utf8]{inputenc}
\usepackage{amsmath}
\usepackage{amssymb}
\usepackage{graphicx}
\graphicspath{{./images/}}
\usepackage[hidelinks]{hyperref}
\usepackage[utf8]{inputenc}
\usepackage[english]{babel}
\usepackage{siunitx}
\usepackage{MnSymbol} 

\usepackage{amsmath}
\usepackage{mathtools}
\newtagform{supplementary}[S.]()
\counterwithin*{equation}{part}
\usepackage{amsfonts}
\usepackage{amssymb}
\usepackage{gensymb}
\usepackage{graphicx}
\usepackage{xcolor}
\usepackage{cleveref}
\usepackage{authblk}
\usepackage{placeins}

\usepackage[style=nature]{biblatex}
\usepackage{float}
\usepackage{xr}

\usepackage{geometry}
\usepackage{times}
\usepackage{longtable}
\usepackage{comment}
\usepackage[textsize=tiny]{todonotes}

\usepackage{amsthm}

\title{Mitochondrial mechanics nucleates axonal jamming and swelling}

\author[1]{Patrick S. Noerr} 
\author[2]{Ahmed A. Abushawish}
\author[2]{Gulcin Pekkurnaz} 
\author[1,3]{Padmini Rangamani\thanks{prangamani@health.ucsd.edu}} 
\affil[1]{Department of Pharmacology, School of Medicine, University of California San Diego}
\affil[2]{Department of Neurobiology, School of Biological Sciences, University of California San Diego}
\affil[3]{Department of Mechanical and Aerospace Engineering, University of California San Diego.}
\begin{document}


\maketitle

\section{Abstract}

Neuronal function requires precise spatial organization of mitochondria to meet localized energetic demand.
However, the physical constraints governing mitochondrial transport in axons remain poorly defined. 
Bidirectional motor-driven trafficking inherently introduces the potential for collisions, but the implications of these interactions for transport failure and structural damage are not understood. 
Here, we develop an agent-based model that couples mitochondrial motility, morphology, and lifecycle dynamics to a deformable axonal boundary. 
We show that mitochondrial traffic jams emerge from a force balance between active propulsion and steric interactions, and that their severity is governed by organelle shape and mechanical properties. Elongated, mechanically rigid mitochondria remain aligned and are transported rapidly, whereas flexible, low-aspect-ratio mitochondria are prone to jamming and accumulation. 
Incorporating fission and fusion dynamics reveals that fission amplifies transport disruption by generating collision-prone populations, while fusion restores transport by producing anisotropic structures that navigate crowded environments more efficiently. 
Importantly, we find that sustained jamming generates mechanical stress on the axonal membrane, leading to deformation and swelling. Together, these results establish a physical framework linking mitochondrial dynamics to axonal integrity and provide testable predictions for how dysregulated fission-fusion balance can drive transport failure and structural pathology in neurons.

\section{Significance}

Axonal deformation is implicated in myriad neurodegenerative conditions. Mitochondrial transport disruption is inextricably linked to axonal deformation and disease progression. 
Mechanistic understanding of the interplay between mitochondrial transport and axon stability remains opaque. 
Here, we developed an agent-based model of mitochondrial transport through axons. 
We found that mitochondria, driven toward presynapses for energy supply and toward the soma for repositioning or recycling, can collide, jam, and accumulate within axonal segments. 
The severity of jamming is sensitive to mitochondrial density as well as mechanical and morphological properties. 
Further, we found a balance between lifecycle dynamics including fission and fusion is paramount to maintaining homeostatic transport. 
Lastly, we predict that accumulated mitochondria can deform the axonal membrane, thereby elucidating a direct mechanical link between mitochondrial transport disruption and axonal deformation.  

\section{Keywords}

agent-based modeling, mitochondria, axon mechanics, motor-driven transport, neurons, active matter

\newpage

\section{Introduction}



Neurons operate under exceptional energetic constraints imposed by their morphology and function \cite{clarke1999basic,simpson2007supply}.
Their polarized architecture enables compartmentalized information processing, but also necessitates precise spatial coordination of energy supply \cite{donato2019neuronal}.
Synaptic activity, including the integrating of neuronal signals and the sustained transmission of action potentials, is energetically expensive \cite{harris2012synaptic,li2022energy}.
Action potentials are initiated at the ion channel-dense axon initial segment proximal to the soma and propagate along extended axonal arbors in a tightly coordinated spatiotemporal manner \cite{kole2008action,bender2009axon,sheffield2011slow,sasaki2011action,bakkum2013tracking}.
The combination of axonal length, reaching up to a meter in humans, and energy demand of neuronal function necessitates a dependable, dynamic, and plentiful source of local ATP production. 
Roughly ninety-three percent of the energy fueling synaptic activity is garnered from oxidative phosphorylation in mitochondria \cite{sokoloff1960metabolism}.


The location and distribution of mitochondria in neurons have been the subject of extensive studies in recent years \cite{schwarz2013mitochondrial,lin2015regulation,pekkurnaz2022mitochondrial,duarte2023mitochondria}. Seminal experiments uncovered several key relationships between mitochondrial and neuronal function.
Mitochondrial dynamics are modulated by synapse proximity and activity \cite{obashi2013regulation}. 
Increased activity upregulates mitochondrial ATP synthesis at synaptic sites \cite{rangaraju2014activity} through local calcium signaling \cite{ghosh2025synapses}. Together, these findings support an emerging feedback relationship linking mitochondrial morphology, transport, and function, synaptic plasticity, and cytoskeletal coupling \cite{rangaraju2019spatially}.  

\begin{figure*}[tbhp]
    \centering
    \includegraphics[width=\linewidth]
    {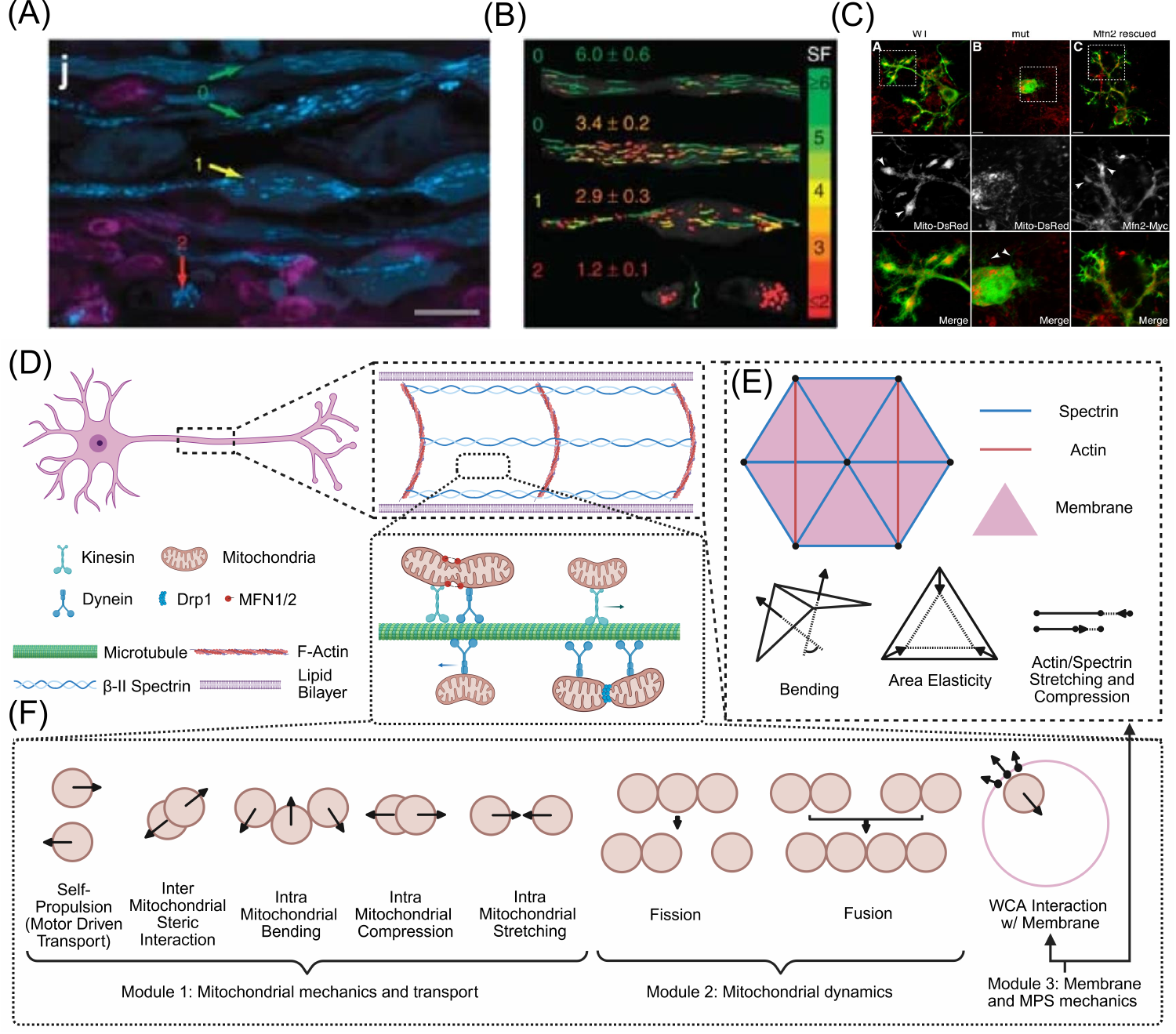}
    \caption{\textbf{Motivation and model schematic of mitochondrial transport through neuronal axon}. (A) Select examples of axonal swelling from the literature. Axonal membrane in multiple sclerosis (MS) mouse models display various morphologies from uniform cylinders to swellings to rupture. (B) Mitochondrial morphologies corresponding to these axonal states with elongated organelles more present in cylindrical axons and isotropic organelles in swelling and ruptured compartments. (A) and (B) are reproduced  from \cite{nikic2011reversible}; permission pending. (C) Mitochondrial lifecycle dynamics influence their distribution in Purkinje dendritic arbors (reproduced from \cite{chen2007mitochondrial}; permission pending). (D) Cartoon of a motor neuron depicting the key components including the lipid bilayer, actin rings and spectrin filaments comprising the membrane periodic skeleton (MPS), microtubules, molecular motors kinesin and dynein, mitochondria bound to motors via adapter proteins, and proteins mediating fusion and fission dynamics. (E) The axon is discretized as a triangular lattice with membrane/cytoskeletal composite forces including bending, area elasticity, and stretch/compression of actin and spectrin in our model. (F) Mitochondria are modeled as soft spheres or bead-spring chains with steric interactions with membrane and other mitochondria, self-propulsion, and fission/fusion dynamics.}
    \label{fig:Schematic}
    \end{figure*}

Mitochondrial transport between the soma and the distal presynaptic terminals is driven by motor protein mediated active transport \cite{saxton2012axonal}. 
The Miro-TRAK adapter complex facilitates the binding of  kinesin and dynein family motor proteins to mitochondria \cite{kruppa2021motor}. 
These molecular motors then hydrolyze ATP to ``step" along heterodimeric tubulin comprising microtubules in a unidimensional fashion with kinesin walking toward the plus end and dynein toward the minus end \cite{bhabha2016dynein,deguchi2023direct}. 
In neuronal axons, microtubule bundles in the axon shaft are oriented with their plus end toward presynaptic terminals and their minus end directed toward the nucleus \cite{heidemann1981polarity}. 
Thus, anterograde transport is described by kinesin motors shuttling healthy mitochondria toward presynaptic terminals to power synaptic transmission. 
Conversely, retrograde transport is described by dynein motors shuttling mitochondria toward the nucleus for repositioning or, alternatively, lysosomal degradation and recycling \cite{perlson2010retrograde}.
While this bidirectional transport allows for dynamic positioning and optimization of mitochondrial function in neurons, it is not immune to disruption. 
Indeed, impaired mitochondrial motility has been implicated in many neurodegenerative conditions such as amyotrophic lateral sclerosis, motor neuron disease, Charcot-Marie-Tooth disease, Alzheimer’s disease, Huntington’s disease, and Parkinson’s disease \cite{de2008role}. 
While there are many physiological consequences of obstructed mitochondrial transport \cite{marinkovic2012axonal,puls2003mutant,hafezparast2003mutations,stokin2005axonopathy,de2007familial}, the mechanical causes and consequences of impeded mitochondrial transport remain largely unexplored.
Observations of enriched mitochondrial density  within the 
axon swellings (\Cref{fig:Schematic}A) of mouse spinal cord and cat retinal axons suggest traffic jamming as a possible mechanism of impeded mitochondrial transport.
\cite{nikic2011reversible,greenberg1990irregular}. 

Several experiments have additionally implicated mitochondrial morphology (\Cref{fig:Schematic}B), governed by fusion and fission events comprising mitochondrial lifecycle dynamics, in transport malfunction. 
In rat dorsal root ganglion (DRG) cultures, mutants not expressing Mitofusin 2 (Mfn2), one of the two proteins responsible for the fusing of mitochondrial outer membrane, exhibit inhibited transport rates \cite{misko2010mitofusin}. 
Mouse models lacking dynamin-related protein (Drp1), the predominant protein implicated in the fission of the mitochondrial outer membrane, lacked mitochondria in distal dopamine axonal compartments, while also reducing mitochondrial motility in neuronal cultures \cite{berthet2014loss}. 
Both fission and fusion protein knockdowns produced transport dystrophies in Purkinje cells, where mitochondria were not able to escape the proximal somal region \cite{chen2007mitochondrial,fukumitsu2016mitochondrial} (\Cref{fig:Schematic}C).
Taken together, these observations of mitochondrial transport disruption, lifecycle effects, and axonal deformation warrant a critical investigation into the nature of the mechanical coupling between mitochondria and the axonal membrane.

Capturing the microscale interactions of mitochondria with each other, the cytoskeleton, and the neuronal membrane in axons remains technically challenging due to limitations in spatiotemporal imaging resolution, the difficulty of resolving organelle-cytoskeleton contacts in confined geometries, and the complexity of genetically perturbing and tracking these dynamics across models that recapitulate transport deficits. Together, these constraints limit the ability to directly interrogate the mechanical coupling underlying mitochondrial transport. Therefore, computational models that explore these multiphysics interactions provide a necessary complement to experimental approaches and can reveal
the governing physical principles underlying such complex biophysical processes.
Previous models have considered non-interacting mitochondrial transport through static hierarchical dendritic networks and shown that motile mitochondria produce steady-state densities consistent with experimental observations \cite{donovan2024dendrite}. Another transport model linked axonal transport malfunction to regions of disordered microtubules ``swirls" \cite{kuznetsov2009macroscopic}. Worm-like chain models of mitochondria subject to various point-like force patterns give rise to motility regimes observed in experiments of organelles in disrupted cytoskeletal networks \cite{fernandez2024deciphering}. 
Coarse-grained molecular dynamics models of the membrane periodic skeleton (MPS) have validated axonal structure \cite{zhang2017modeling}, and discretized mechanical membrane models have shown large-scale axonal deformations, resulting from a pearling instability \cite{griswold2025membrane}. However, these approaches do not integrate mitochondrial transport, morphology, lifecycle dynamics, and axonal mechanics within a unified framework. 

In this work, we integrate mitochondrial motility, fission and fusion dynamics, and axonal membrane mechanics to determine how their interactions give rise to mitochondrial jamming and axonal swelling in an agent-based modeling framework. 
Our model shows that elongated mitochondria are less likely to form traffic jams while globular mitochondria are more likely to cause jamming in axons. 
Our simulations predict that mitochondrial fission exacerbates transport disruption by generating collision-prone populations,
while mitochondrial fusion restores transport by promoting anisotropic morphologies. 
Finally, we identify conditions under which mitochondrial jamming generates sufficient mechanical stress to deform the axonal membrane and cause it to swell.
Together, these results establish a biophysical framework for the coupling of mitochondrial transport to axonal morphology, and generate experimentally testable predictions. 

\section{Model Development}
We constructed an agent-based model for mitochondrial transport in axons. 
The main elements of our modeling framework are shown in \Cref{fig:Schematic}D-F.
Our model consists of three modules: mitochondrial transport and mechanics, mitochondrial fission and fusion dynamics, and mechanics of the axonal membrane-MPS composite.

\subsection{Assumptions}

We assume a roughly equal population of anterograde and retrograde mitochondria.
While it is likely that the majority of motile mitochondria in neurons are anterograde \cite{ligon2000movement}, this assumption allows us to focus on the collective behavior of the mechanical interactions of mitochondria.
We do not explicitly model kinesin and dynein walking along microtubules.
Rather, we coarse-grain their action into anterograde and retrograde movement of mitochondria with a prescribed velocity of self-propulsion.
This depiction of mitochondrial motility corresponds to the limit in which motor proteins and microtubules are plentiful \cite{hollenbeck1989distribution,yamada1971ultrastructure}. 
The action of the proteins responsible for mitochondrial fission and fusion is represented using rate constants for fission and fusion in our model, similar to \cite{arkfeld2025whole}.

\subsection{Tracking and counting mitochondria}

For each mitochondrial bead, we keep track of its three-dimensional position, particle ID, and chain ID. 
The three-dimensional position is updated according to a forward Euler scheme of a sum of forces. 
Each bead's particle ID is a unique integer given to the bead at the time it is generated in the simulation.
A mitochondrion is generated by placing the rear bead of an anterograde mitochondria at x = 0 with a random y and z position. 
Then, depending on the length of the mitochondrion, $N_\text{chain}$, subsequent beads are given the same y and z position with x values incrementing by intervals of mitochondrial bead diameter, $2r_{m}$, until there are $N_\text{chain}$ beads in that mitochondrial chain. 
The same process is used for retrograde mitochondria except the initial x value is at L = \SI{3.8}{\micro\meter}
, the length of the axon. Subsequent beads then increment their x value by $-2r_{m}$. 
The chain ID of each particle is initialized at the time of generation and subject to change due to fission and fusion events. The chain ID represents the organelle and the bead ID represents a section of length of that organelle.

\subsection{Module 1: Mitochondrial mechanics and transport}

In neurons, mitochondrial morphologies assume a range of aspect ratios from small spheres to elongated slender bodies \cite{fischer2018morphology}. 
We represent a single mitochondrion as a bead spring chain with harmonic bending and stretching energies as described in Ref. \cite{lisicki2024twist}.
Each mitochondrion interacts with other mitochondria through a short range steric Hookean term to enforce volume exclusion. 
We coarse grain mitochondrial transport through molecular motors walking along microtubules with mitochondria in tow into a self-propulsion velocity assigned to each mitochondrial bead representing the force exerted on the mitochondria by bound motors (\Cref{fig:Schematic}F). 
We now write the equations of motion for mitochondria as 

\begin{align}
    \frac{d\vec r^{i,m(a)}}{dt} &= \frac{1}{\gamma_{m}} \vec F^{i,m(a)} + v_{a}\hat n^{i,m(a)}, \\
    \frac{d\vec r^{i,m(r)}}{dt} &= \frac{1}{\gamma_{m}}\vec F^{i,m(r)} + v_{r}\hat n^{i,m(r)},
\end{align}

\noindent where $\vec r^{i,m(a)}$ and $\vec r^{i,m(r)}$ are the position vectors of anterograde mitochondrial bead \textit{i,m(a)}, and retrograde mitochondrial bead \textit{i,m(r)}, respectively. The orientation of anterograde mitochondria \textit{i}, $\hat{n}^{i,m(a)}$, is $+\hat{x}$ and $-\hat{x}$ for retrograde mitochondria, $\hat{n}^{i,m(r)}$. $v_{a}$ and $v_{r}$ are the velocities of anterograde and retrograde mitochondria, respectively, and $\gamma_{m}$ is the viscous damping coefficient of mitochondria in the axoplasm. The total force on mitochondrial bead \textit{i}, $\vec{F}^{i,m}$, can be expanded in terms of interchain, intrachain, and membrane interactions. Thus, the force acting on a single mitochondrial bead is 

\begin{equation}
    \vec F^{i,m} = \vec F^{i,m}_{\text{bend}} + \vec F^{i,m}_{\text{spring}} + \vec F^{i,m}_{\text{steric}} + \vec F^{i,m}_{\text{WCA}}.
\end{equation}

\noindent Here, $\vec F^{i,m}_{\text{bend}}$ is the force from a quadratic bending energy between collinear bonds in a mitochondrial chain given by


\begin{equation}
    \vec F^{i,m}_{\text{bend}} = - (\vec F^{j,m}_{\text{bend}} + \vec F^{k,m}_{\text{bend}})= -\frac{\partial U_{b}}{\partial \theta_{ijk}} \frac{\partial \theta_{ijk}}{\partial \vec r^{i}} = \frac{k_\text{b,mito} \theta_{ijk}}{4r_{m}} \frac{\partial \theta_{ijk}}{\partial \vec{r}^{i}}, 
\end{equation}

\noindent where $i$ is the central mitochondrial bead, and $j,k$ are neighboring beads in the same chain. $\theta_{ijk}$ is the angle between vectors $\vec{r}^{j} - \vec{r}^{i}$ and $\vec{r}^{k} - \vec{r}^{i}$, defined by $\theta_{ijk} = \arccos{ \left( \hat{t}^{i} \cdot \hat{t}^{i+1} \right)}$, where $\hat{t}^{i} = \frac{\vec{r}^{i} - \vec{r}^{j}}{|\vec{r}^{i} - \vec{r}^{j}|}$ and $\hat{t}^{i+1} = \frac{\vec{r}^{k} - \vec{r}^{i}}{|\vec{r}^{k} - \vec{r}^{i}|}$.
The bending stiffness of the chain is $k_\text{b,mito}$. 

\noindent $\vec F^{i,m}_{\text{spring}}$ is a pairwise Hookean spring force between two beads in a chain given by

\begin{equation}
    \vec F^{i,m}_{\text{spring}} = \sum_{j = i+1,i-1} k_\text{s,mito}\left( |\vec{r}^{ij}| - 2r_{m} \right)\hat{r}^{ij},
\end{equation}

\noindent where $\vec{r}^{ij}$ is the separation vector of beads $i$ and $j$, $r_{m}$ is the radius of a mitochondrial bead, and $k_\text{s,mito}$ is the spring constant.
The steric interaction, $\vec F^{i,m}_{\text{steric}}$, is a Hookean interaction modified with a Heaviside function between non-adjacent beads \cite{rickman2019effects} written as

\begin{equation}
    \vec F^{i,m}_{\text{steric}} = \sum_{j \neq i+1,i-1}^{N} k_{M}\left( |\vec{r}^{ij}| - 2r_{m} \right)\Theta\left( 2r_{m}-|\vec{r}^{ij}| \right)\hat{r}^{ij},
\end{equation}

\noindent where $\vec{r}^{ij}$ is the separation vector of beads $i$ and $j$, $k_{M}$ is the steric parameter, $r_{m}$ is the radius of a mitochondrial bead, and $\Theta(x) = 1$ when $x \geq 0$ and $\Theta(x) = 0$ when $x < 0$. 

\noindent Lastly, the interaction between the axonal membrane and motile mitochondria in the axoplasm is modeled by a WCA potential \cite{shendruk2014coarse},
\begin{equation}
    \vec F^{i,m}_{\text{WCA}} = -\frac{24\phi_{0}}{r_{m}}\sum_{j \in N_{M}}\left[2\left(\frac{|\vec{r}^{ij}|}{r_{m}}\right)^{-13}-\left(\frac{|\vec{r}^{ij}|}{r_{m}}\right)^{-7}\right]\hat{r}^{ij}, \qquad |\vec r^{ij}| < 2^{1/6}r_{m},
\end{equation}
where $r_{m}$ is the radius of the mitochondrial bead, $N_{a}$ are the set of membrane nodes, $\vec{r}^{ij}$ is the separation vector between mitochondrial bead $i$ and membrane node $j$, and $\phi_{0}$ is the strength of the WCA interaction. Equations (1) - (7) are non-dimensionalized for numerical implementation with the length scale $r^{*}=\frac{r}{r_{0,a}}$ and time scale $t^{*} = \frac{v_{0}}{r_{0,a}}t$, where $r_{0,a}$ and $v_{0}$ are the initial radius of the axon and average velocity of mitochondria as described in \Cref{tab:params}.

\subsection{Module 2: Mitochondrial lifecycle dynamics}

Mitochondria dynamically undergo fission and fusion in response to myriad cellular cues \cite{van2013mechanisms}. 
We implement this feature in our model by including a dynamic chain ID (\Cref{fig:Schematic}F). 
To implement fission, at each timestep, we loop over each mitochondria and generate a random number $0 \leq p_{i} < 1$ from a uniform distribution using numpy.random.random for mitochondria $i$. 
If the product of the timestep $\Delta t$ and the fission rate constant $k_\text{fission}$ is greater than or equal to $p_{i}$, fission occurs at a random bead within the chain using numpy.random.choice. The resulting beads are given a new chain ID indicating that they are no longer a part of the original mitochondria prior to the fission event.

To implement fusion, at each timestep, we loop over every pair of mitochondrial beads that have at most one neighbor.
Fusion occurs if the following three conditions are met: (a) We generate a random number $0 \leq p_{i} < 1$ from a uniform distribution using numpy.random.random and the product of the timestep $\Delta t$ and the fusion rate constant $k_\text{fusion}$ is greater than or equal to $p_{i}$.
(b) The distance between beads is less than or equal to $r_\text{fusion}$.
This is imposed so that neighboring mitochondria must be proximal to one another to fuse.
(c) The difference in orientation of the mitochondria is less than or equal to $\theta_\text{cutoff}$. 
If all three of these conditions are met, the chain IDs of the two fusing mitochondria are made the same, indicating all beads of these mitochondria are part of the same organelle. 
If the fused mitochondria were both anterograde (retrograde), the fused mitochondria is anterograde (retrograde). 
If one of the mitochondria was anterograde and the other was retrograde, the fused mitochondria becomes anterograde or retrograde at random using numpy.random.choice.

\subsection{Module 3: Mechanics of the axonal membrane and the membrane periodic skeleton}
In this module, we incorporate the mechanics of the axonal membrane and the membrane periodic skeleton.
The central structural components of the axon are the lipid bilayer comprising the plasma membrane and the MPS, which is an arrangement of filamentous actin rings spaced 180 nm apart and connected by spectrin filaments \cite{xu2013actin}. 
We model the axonal membrane-MPS in simulations as a composite of these components (\Cref{fig:Schematic}E). 
The forces exerted on the axon at node $i$, $\vec F^{i,a}$, can be expanded in terms of contributions from bending, stretching and compression, cytoskeletal deformations, and interactions with mitochondria, written as
\begin{equation}
    \vec F^{i,a} = \vec F^{i,a}_{\text{membrane bending}} + \vec F^{i,a}_{\text{area elasticity}} + \vec F^{i,a}_{\text{cytoskeleton}} + \vec F^{i,a}_{\text{WCA}}.
\end{equation}
The bending force given by 
\begin{equation}
    \vec F^{q}_{\text{membrane bending}} =  2k_{b}\frac{||e_{ij}||}{A_{ij}}\sin\theta_{ij} \nabla_{q} \theta_{ij},
\end{equation}
represents the forces on nodes $q \in \{i,j,k,l\}$, where nodes $i,j,k$ and $i,j,l$ form triangles sharing the hinge formed by segment $i,j$, due to the pure bending modes of a quadratic bending energy functional with bending stiffness $k_{b}$ described in \cite{wardetzky2007discrete}. The sum of bending forces on axon node \textit{i} is, therefore, the sum of all bending forces in which node \textit{i} is one of the four vertices, \textit{q}.
The area elasticity force,
\begin{equation}
    \vec F^{i,a}_{\text{area elasticity}} = -\frac{k_{AE}}{A_{0}}\left(\frac{A}{A_{0}}-1\right)\sum_{f \in F_{i}} \nabla^{i}A_{f},
\end{equation}
represents the tendency for the membrane to preserve its surface area, $A_{0}$, with strength $k_{AE}$, where $F_{i}$ are all the triangular mesh faces that contain node \textit{i}. $\nabla^{i}A_{f} = \frac{1}{2}\hat{n}_{f}\times \left(\vec{r}^{k}-\vec{r}^{j}\right)$, where $\hat{n}_{f} = \frac{\left(\vec{r}^{j}-\vec{r}^{i}\right) \times \left(\vec{r}^{k}-\vec{r}^{i}\right)}{|\left(\vec{r}^{j}-\vec{r}^{i}\right) \times \left(\vec{r}^{k}-\vec{r}^{i}\right)|}$ as described in \cite{runser2024simucell3d}.
The tendency of the cytoskeletal filaments comprising the MPS to resist deformation is modeled by Hookean spring interactions,
\begin{gather}
    \vec F^{i,a}_{\text{cytoskeleton}} = \vec F^{i,a}_{\text{spectrin}} + F^{i,a}_{\text{actin}}, \\
    \nonumber \text{where} \\
    \vec F^{i,a}_{\text{spectrin}} = -k_{S} \sum_{j \in N_{S}}\left(|\vec{r}^{ij}|-d_{0,S}\right)\hat{r}^{ij} \\
    \nonumber \text{and} \\
    \vec F^{i,a}_{\text{actin}} = -k_{A} \sum_{j \in N_{A}}\left(|\vec{r}^{ij}|-d_{0,A}\right)\hat{r}^{ij}. \\ \nonumber 
\end{gather}
$N_{S}$ are neighboring nodes sharing a spectrin bond, $N_{A}$ are neighboring nodes sharing an actin bond. $k_{S}$ and $k_{A}$ are the stiffness of spectrin and actin filaments with rest lengths $d_{0,S}$ and $d_{0,A}$, respectively. Lastly, we include the pairwise interaction between membrane node \textit{i} and mitochondria bead \textit{j} in the axoplasm with

\begin{equation}
    \vec F^{i,a}_{\text{WCA}} = -\vec F^{j,m}_{\text{WCA}} = -\frac{24\phi_{0}}{r_{m}}\sum_{j \in N_{m}}\left[2\left(\frac{|\vec{r}^{ij}|}{r_{m}}\right)^{-13}-\left(\frac{|\vec{r}^{ij}|}{r_{m}}\right)^{-7}\right]\hat{r}^{ij}, \qquad |\vec r^{ij}| < 2^{1/6}r_{m}.
\end{equation}

\noindent Axonal nodes are updated with a forward Euler scheme in time according to the overdamped equation,

\begin{equation}
    \frac{d\vec r^{i,a}}{dt} = \frac{1}{\gamma_{a}} \vec F^{i,a}.
\end{equation}

\noindent Equations (8) - (15) are non-dimensionalized for numerical implementation with the length scale $r^{*}=\frac{r}{r_{0,a}}$ and time scale $t^{*} = \frac{v_{0}}{r_{0,a}}t$, where $r_{0,a}$ and $v_{0}$ are the initial radius of the axon and average velocity of mitochondria as described in \Cref{tab:params}.

\begin{table*}[ht!]
    \centering
    \caption{Model Parameters}
    \begin{tabular}{ | m{3.75cm} | m{3.75cm}| m{3.75cm} | m{3.75cm} |} 
        \hline
        Parameter name & Value(s) & Reference & Parameter meaning \\
        \hline
        $r_{0,a}$ & $500  \ \text{nm}$ & \cite{perge2012axons} & Initial radius of axon\\
        \hline
        $l_{0}$ & $38 \ \text{nm}$ & - & cylindrical lattice (spectrin) rest length\\
        \hline
        $l_{a}$ & $65 \ \text{nm}$ & - & actin rest length\\
        \hline
        $r_{m}$ & $125 \ \text{nm}$ & \cite{teixeira2024super} & mitochondrial radius\\
        \hline
        $v_{0}$ & \SI{0.5}{\micro\meter}$\text{s}^{-1}$ & \cite{kang2008docking} & Average velocity of mitochondria \\
        \hline
        $v_{a}$ & \SI{0.5}{\micro\meter}$\text{s}^{-1}$ & \cite{kang2008docking} & Anterograde mitochondria velocity \\
        \hline
        $v_{r}$ & \SI{0.5}{\micro\meter}$\text{s}^{-1}$ & \cite{kang2008docking} & Retrograde mitochondria velocity \\
        \hline
        $\phi$ & $0.3\phi_\text{max}$ - $0.9\phi_\text{max}$ & - & Mitochondrial density \\
        \hline
        $N_\text{chain}$ & 1-5 & - & Number of beads per mitochondria \\
        \hline
        $\gamma_{a}$ & $1 \ \text{kgs}^{-1}$
        & - & viscous damping coefficient of axon \\
        \hline
        $\gamma_{m}$ & $1 \ \text{kgs}^{-1}$ & - & mitochondrial drag coefficient \\
        \hline
        $k_{b}$ & $5 \times 10^{-19} \ \text{J}$ & \cite{bochicchio2016membrane,simson1998membrane,lang2017axonal} & Membrane bending rigidity \\
        \hline    
        $k_{AE}$ & $10^{-15} \ \text{J}$ & \cite{runser2024simucell3d} & Area elasticity modulus \\
        \hline
        $k_{S}$ & $2.06 \times 10^{-4}  \ \text{Nm}^{-1}$ & \cite{hossain2024mechanical}  & Spectrin spring coefficient \\
        \hline
        $k_{A}$ & $0.26 \ \text{Nm}^{-1}$ & \cite{zhang2017modeling} & Actin spring coefficient \\
        \hline
        $k_{M}$ & $40 \ \text{Nm}^{-1}$ & - & Mito-mito spring coefficient \\
        \hline
        $\phi_{0}$ & ~$5 \times 10^{-15} \ \text{J}$ & - & Mito-membrane WCA prefactor \\
        \hline
        $k_\text{b,mito}$ & $10^{-21} - 10^{-19} \ \text{Nm}^{2}$ & - & Mitochondrial bending rigidity \\
        \hline
        $k_\text{s,mito}$ & $100 \ \text{Nm}^{-1}$  & - & Mitochondrial spring coefficient \\
        \hline
        $k_\text{fission}$ & $.01 - 100 \ \text{s}^{-1}$ & \cite{arkfeld2025whole} & Mitochondrial fission rate \\
        \hline
        $k_\text{fusion}$ & $.01 - 100 \ \text{s}^{-1}$ & \cite{arkfeld2025whole} & Mitochondrial fusion rate \\
        \hline
        $r_\text{fusion}$ & $1.1 \times 2r_{m}$ & - & Threshold fusion distance \\
        \hline
        $\theta_\text{cutoff}$ & $30^{\circ}$ & - & Threshold orientation difference \\
        \hline
    \end{tabular}
    \label{tab:params}
\end{table*}

\subsection{Simulation Protocol}
Initially, two anterograde and retrograde mitochondria are generated on either side of the \SI{3.8}{\micro\meter} long cylinder representing a portion of the axon shaft. Every $10 \times N_\text{chain}$ subsequent timesteps, which is implemented so $\frac{dN_\text{beads}}{dt} =$ const across chain lengths, anterograde and retrograde mitochondria are symmetrically generated until the number of mitochondria reaches $N_\text{mito}$, where $N_\text{mito}$ is calculated according to a specified density $\phi = x\phi_\text{max}$ where $\phi_\text{max}$ is the maximum filling fraction of hexagonally closed packed spheres in a given volume ($\phi_\text{max} \approx .7405$ \cite{hales2017formal}). Mitochondria that escape the axon are dynamically removed from their respective arrays.
The aforementioned equations and operations are implemented in a custom Python script with parameters stated in Table \ref{tab:params}, biophysically visualized in the schematic shown in \Cref{fig:Schematic}D-F, and described algorithmically in \Cref{fig:Algorithm}. 
The code to run all of the simulations, along with the analysis scripts and sample trajectories can be found here \cite{patricknoerr_2026_19657215}.

\section{Results}

To establish the governing principles of mitochondrial transport in axons, we first examined whether bidirectional mitochondrial transport is sufficient to generate transport disruptions in a confined axonal geometry. 
Modeling mitochondria as self-propelled bead–spring chains, we observed a density-dependent jamming that is characterized by a pronounced reduction in ensemble velocity. 
This behavior emerges from a force balance between active self-propulsion and steric interactions arising from collisions between anterograde and retrograde populations. 
To quantify these disruptions, we defined a relief time as the time required to recover unimpeded velocity, and computed nematic order and shape factor to track orientational and morphological changes of the mitochondria.
We find that mitochondrial mechanics strongly influence jamming: elongated, rigid mitochondria maintain alignment and are less prone to arrest, whereas low–aspect ratio, deformable mitochondria form persistent accumulations. 
Incorporating lifecycle dynamics shows that fission generates collision-prone populations that exacerbate jamming, while fusion promotes elongation and facilitates transport. 
Finally, sustained jamming produces mechanical stresses on the axonal boundary, leading to membrane deformation. 
Together, these results link mitochondrial morphology and mechanics to transport disruption and axonal swelling.

\subsection*{Bidirectional mitochondrial transport generates density-dependent jamming in a confined axonal geometry}

\begin{figure*}[tbhp]
    \centering
    \includegraphics[width=\linewidth]
    {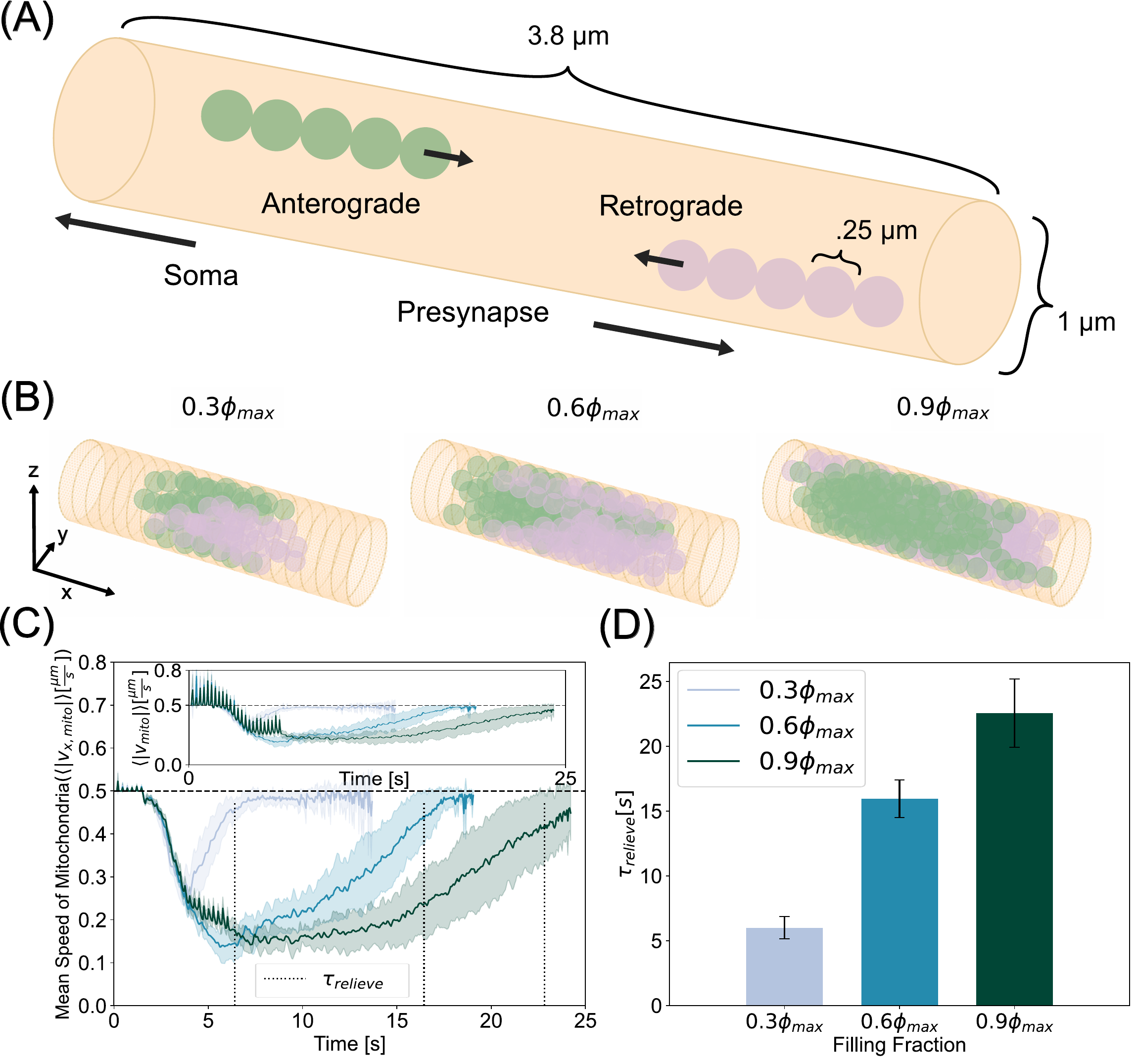}
    \caption{\textbf{Mitochondria modeled as bidirectionally translated bead-spring chains exhibit jamming in cylindrical axons}. (A) Schematic indicating bead spring chain mitochondria undergoing anterograde (green) and retrograde (plum) transport. (B) Simulation snapshots at the time of most severe jamming for low, moderate, and high mitochondrial densities. This time of most severe jamming is different for each mitochondrial density. (C) Time series of the average speed along the axonal long axis of mitochondria for three different densities. Ensemble averages of absolute velocities show a rapid drop after retrograde and anterograde populations collide. At higher densities, jammed states take longer to be relieved. Inset shows full velocity profile, where early times exhibit fluctuations due to steric interactions giving rise to large radial displacements while mitochondrial chains are being generated. Dashed vertical lines indicate relief time, $\tau_\text{relieve}$. (D) Time taken for ensemble speeds to recover 95\% of the mitochondrial self-propulsion speed. While sparse mitochondrial collisions can be alleviated quickly, dense systems take longer to overcome impingement. Curves in (C) and bar heights in (D) represent the mean of ten simulations per density. The shaded region in (C) and the errorbars in (D) represent the standard deviation of the same data.}
    \label{fig:fig3}
    \end{figure*}

To investigate the nature of bidirectional mitochondrial transport in axons, we constructed a cylinder comprised of a triangular mesh and allowed the system to evolve in time as described in Model Development. 
We first focused on mitochondrial dynamics in a cylindrical geometry by imposing a fixed axon, i.e., $ \frac{d\vec{r}^{i,a}}{dt} = 0 $, regardless of $\vec{F}^{i,a}$. 
To simulate the different factors that may affect energy demand in the axon including local activity, available motor concentration, and axon size, we varied the density of mitochondria in our simulations. 
By varying the density, we effectively varied the number of mitochondria (or the volume fraction of mitochondria occupied in the axon). 
We tested three densities, $\phi = 0.3\phi_\text{max}, 0.6\phi_\text{max}, 0.9\phi_\text{max}$, corresponding to 81, 162, and 243 beads, respectively, where $\phi_\text{max} \approx .7405$ is the maximum volume fraction for a closed hexagonal packing of spheres as stated in Model Development. 
These densities represent a sparse, intermediate, and dense packing of mitochondria, respectively. 
We simulated elongated mitochondria of aspect ratio 5 ($N_\text{chain} = 5$). 
In the simulations, mitochondrial beads have a diameter of 250 nm, mitochondrial chain rest lengths are $1.25 \ \mu \text{m}$, and anterograde and retrograde velocities are set to \SI{0.5}{\micro\meter}$\text{s}^{-1}$ (\Cref{fig:fig3}A). 

Our simulations were set up such that anterograde and retrograde populations meet near the middle of the axon where self-propulsion and steric interactions between the mitochondria compete in a force balance. Self-propulsion pushes the mitochondria toward the opposite end of the axon from where they were generated and steric interactions, pushing mitochondria away from each other, are largely pushing mitochondria away from the opposite end of the axon where they were generated.
We hypothesized that this competition would slow down the mitochondrial velocity through the axon, resulting in a disruption of mitochondrial transport (\Cref{fig:fig3}B, Movie 1, Movie 2). 
To quantify this disruption, we tracked the ensemble averaged speeds of all mitochondrial beads, including both anterograde and retrograde populations, in time and plot the average speeds of ten simulations with a shaded region representing the standard deviation of ten simulations (\Cref{fig:fig3}C - inset). At early times, we saw fluctuations due to large steric forces as mitochondria are being generated. 
Since we are predominantly interested in motility along the long axis of the axon, from here onwards, we plot the \textit{x}-component of the ensemble speeds (\Cref{fig:fig3}C).

Our simulations showed that the mean speed of mitochondria in the axons depends on the density of mitochondria, as expected. 
We also observed a decrease in the mean speed of mitochondria over time for all densities, indicating that mitochondria were jammed for finite time periods in the axon. We calculated the time required to relieve this jamming from the simulations as the time it took for the mean speed to recover to baseline $\left( \langle |v_\text{x,mito}|\rangle = .95|v_{0}| \right)$
We found that while all mitochondrial densities exhibited a time of transport hindrance, higher densities experienced a larger magnitude and longer period of disruption.
We defined the time it takes for mitochondrial ensembles to recover 95\% of their imputed self-propulsion velocity as the relief time, denoted by $\tau_\text{relieve}$. 
Higher mitochondrial densities took longer to relieve the emergent transport disruption resulting from the force balance of propulsion and repulsion (\Cref{fig:fig3}D). 
Thus, simulations from our model predict that mitochondria can temporarily accumulate into traffic jams in axons, reducing their mean velocities, and increasing their residence time. 
Elevated mitochondrial densities prolong residence times.
This trend is consistent with experimental observations of mouse models of Alzheimer's where mutants displaying more mitochondria per neuron were accompanied by reduced motility \cite{calkins2011impaired}.

\subsection*{Mitochondrial mechanics affects jamming severity}
\begin{figure*}[tbhp]
    \centering
    \includegraphics[width=.9\linewidth]
    {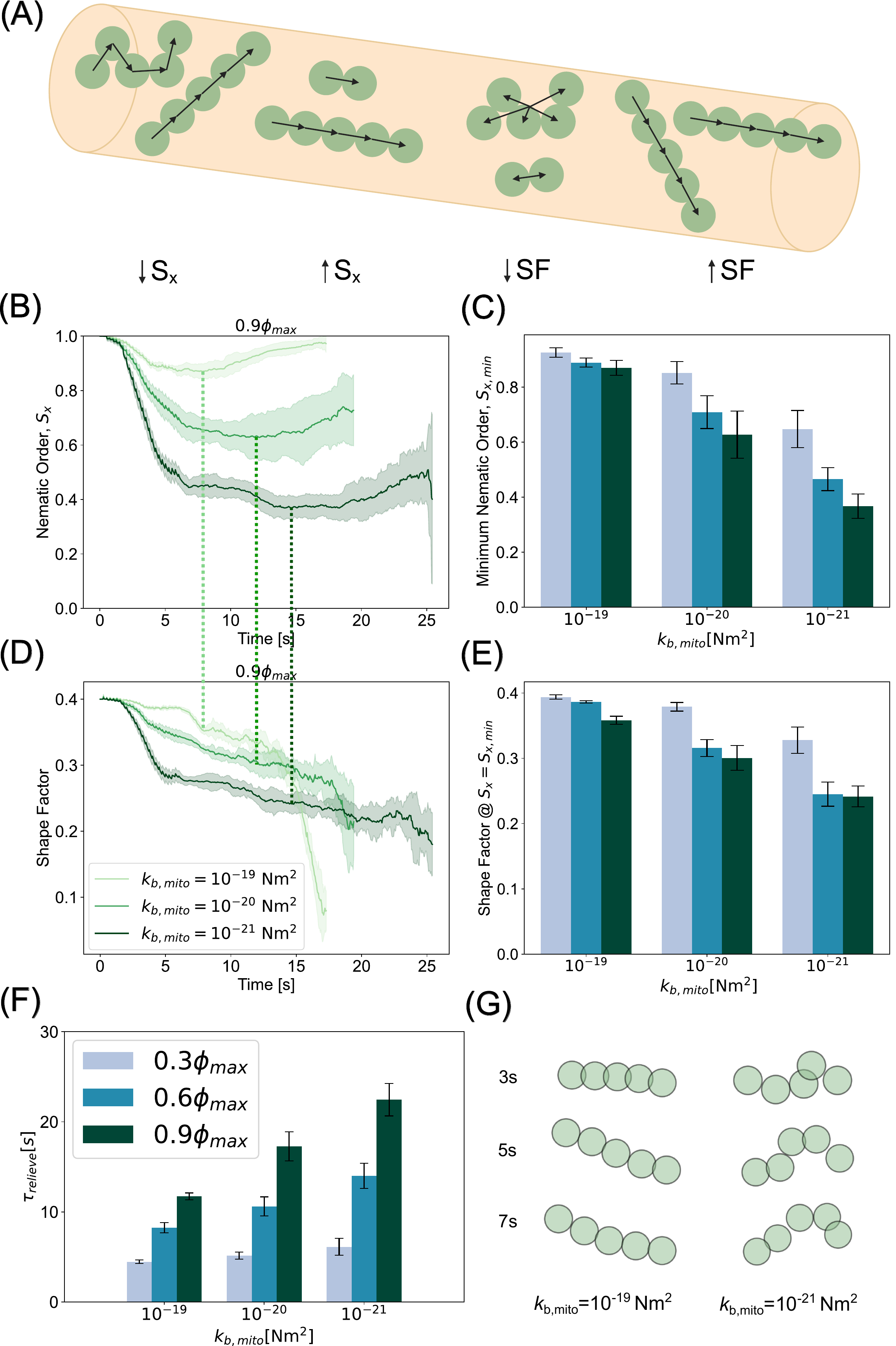}
    \caption{(Caption next page.)}
    \label{fig:fig4}
    \end{figure*}

Mitochondria are subject to several forces from motor proteins, the cytoskeleton, and interactions with the cytosolic environment \cite{bartolak2017regulation,schiavon2020actin}.  
As a result of these forces, mitochondria can deform and resist both compressive and tensile forces \cite{feng2018mechanical}.
Previous studies have calculated that the persistence length of mitochondria in melanophores is on the order of several microns, indicating a bending rigidity on the order of $10^{-26} \ \text{Nm}^{2}$ \cite{fernandez2022morphological}. It is entirely possible that, in axons, the measured bending rigidity would appear much higher due to the combined effect of increased confinement and a more ordered cytoskeleton. 
However, the exact value of this parameter has not yet been measured \textit{in situ}.
Here, we varied mitochondrial bending rigidity between $10^{-21} \ \text{Nm}^{2}$ and $10^{-19} \ \text{Nm}^{2}$ to explore the effect of bending rigidity on jamming dynamics, while keeping the chain length constant.
Specifically, we investigated how the  configurational and morphological traits of elongated mitochondria $\left( N_\text{chain}=5\right)$ affect their propensity to jam in axons. 
To characterize the configurational characteristics of mitochondria, we computed the nematic order parameter along the long axis of the channel, $S_{x} \equiv \langle \frac{3\cos^{2}\theta-1}{2} \rangle$, where $\theta$ is the angle between the separation vector of two neighboring beads in a chain, $\vec{r}^{ij} = \vec{r}^{j} - \vec{r}^{i}$ and the x director, $\overleftrightarrow x$ (\Cref{fig:fig4}A - left). 
The nematic order parameter represents how well a distribution of orientations can be described by a single axis. 
A value of 1 represents perfect angular consistency among all agents while a value of 0 represents an isotropic configuration (or no preferred orientation). 
Our simulations showed that only the stiffest mitochondria, $k_\text{b,mito} = 10^{-19} \ \text{Nm}^{2}$, fully recovered a nematic order of close to 1, indicating that they underwent only a slight deformation as they were transported through the axon (Movie 3). 
Mitochondria with intermediate stiffness, $k_\text{b,mito} = 10^{-20} \ \text{Nm}^{2}$, exhibited a moderate decrease in nematic order during collision, followed by a partial recovery of alignment. 
Easily deformable mitochondria, with $k_\text{b,mito} = 10^{-21} \ \text{Nm}^{2}$, demonstrated a dramatic decrease in nematic order during anterograde-retrograde collision.
In this case, the nematic order remained low for the duration of the simulation (\Cref{fig:fig4}B).
\addtocounter{figure}{-1}
\begin{figure*}[t!]
    \caption{\textbf{Bending stiffness of mitochondria regulates severity of jamming.} (A) Schematic demonstrating the  order parameters used to analyze characteristic mitochondrial morphologies including the nematic order parameter with respect to the long axis of the channel - $\text{S}_{\text{x}}$ and the shape factor - SF. (B) Nematic order as a function of time at the highest density, $\phi = 0.9\phi_\text{max}$, shows that stiffer filaments experience less buckling and recover overwhelmingly parallel orientations than their more flexible counterparts. Time of minimum nematic order indicated by overlaid dashed lines of color consistent with bending rigidity. (C) Minimum nematic order indicates persistent orientation parallel to axonal segment at higher rigidities for all densities. (D) Shape factor metric reveals morphology of stiffer mitochondria remains largely elongated and anisotropic while more flexible filaments become globular during collisions. (E) Shape factor at the time of minimum nematic order suggests rigid mitochondria retain anisotropic morphology through jamming event. (F) While all flexibilities require approximately the same amount of time to relieve jamming at low density, rigid mitochondria require significantly less time to restore unimpeded speeds at moderate and high densities. (G) Characteristic snapshots of rigid mitochondria (left) and flexible mitochondria (right) confirm morphologies suggested by analyses. Plots and bar height in (B-E) represents mean values of ten simulations per parameter set. The corresponding shaded regions and error bars represent the standard deviation of the same data.}
    \end{figure*}
Thus, our simulations show that in the limit of infinite stiffness, mitochondria behave as an active nematic, while in the limit of infinite compliance, mitochondria behave as tangled globular polymers.
We then calculated the minimum nematic order during mitochondrial collisions as a function of bending stiffness and mitochondrial density. We found that rigid mitochondria retained a nematic order close to one across densities. 
Flexible mitochondria lost angular consistency during collisions, especially at higher density (\Cref{fig:fig4}C). 
Overall, we found that the nematic order depends on the bending stiffness of the mitochondria, with more flexible mitochondria exhibiting greater sensitivity to density.

To further quantify the morphological attributes of mitochondria with different bending stiffnesses, we calculated the shape factor, $\text{SF} = \frac{1}{N_\text{chains}}\sum_{i}^{N_\text{chains}}\left(\frac{1}{N^{i}_\text{chain}}^{2}\right)\sum_{j}^{N_\text{chain}} \left(\frac{|\vec{r}^{j}-\vec{r}^{i}_{COM}|}{2r_{m}} \right)^{2}$, where $\vec{r}^{i}_{COM}$ is the center of mass of chain \textit{i}.
A larger value of the shape factor indicates a higher aspect ratio with an anisotropic morphology and a lower value indicates a more globular isotropic morphology (\Cref{fig:fig4}A - right).
The initial shape factor of an unperturbed chain of length 5 is 0.4. 
We observed that stiffer mitochondria retained a shape factor close to that of an unperturbed chain throughout the transport through the axon while the flexible mitochondria quickly crumpled and assumed a more compact morphology at the onset of collision (\Cref{fig:fig4}D). 
To understand the relationship between the minimum nematic order, which is associated with increased collisions, and mitochondrial morphology, we plotted the shape factor at the time of minimum nematic order.
We found that at any given mitochondrial density, flexibility promotes disorder while rigidity stabilizes orientation and morphology, especially in the sparse regime (\Cref{fig:fig4}C,E).
Taken together, these analyses provide a robust framework for how the mechanical properties of mitochondria can affect orientation and configuration.
In the rigid limit, mitochondria are persistently oriented and elongated, while in the flexible limit, they form globular bundled structures.
We next analyzed the effect of  mitochondrial bending stiffness on transport dynamics.
Our simulations reveal that stiffer mitochondria were able to relieve traffic jamming easier than their more flexible counterparts (\Cref{fig:fig4}F). 
This is because rigid mitochondria behave like active nematic filaments, which are known to exhibit antiparallel laning \cite{memarian2021active}, while flexible mitochondria lose their anisotropic morphology due to collision-induced buckling. 
This trend holds across densities, with the difference between stiffness decreasing at sparse densities 
(\Cref{fig:Fig4_SI}).
Consistent with this rationale, characteristic snapshots of individual chain morphologies near the initial collision time (\Cref{fig:fig4}G) confirm that stiffer mitochondria present an effectively lower cross sectional area for collision than their flexible counterparts. 
Thus, we predict any influence that would tend to alter mitochondrial morphology from straight, slender chains, including active cytoskeletal forces and induced swelling or pearling, will put mitochondria at greater risk for jamming-induced accumulation.

\subsection*{Mitochondrial morphology potently influences jamming}
\begin{figure*}[ht!]
    \centering
    \includegraphics[width=\linewidth]
    {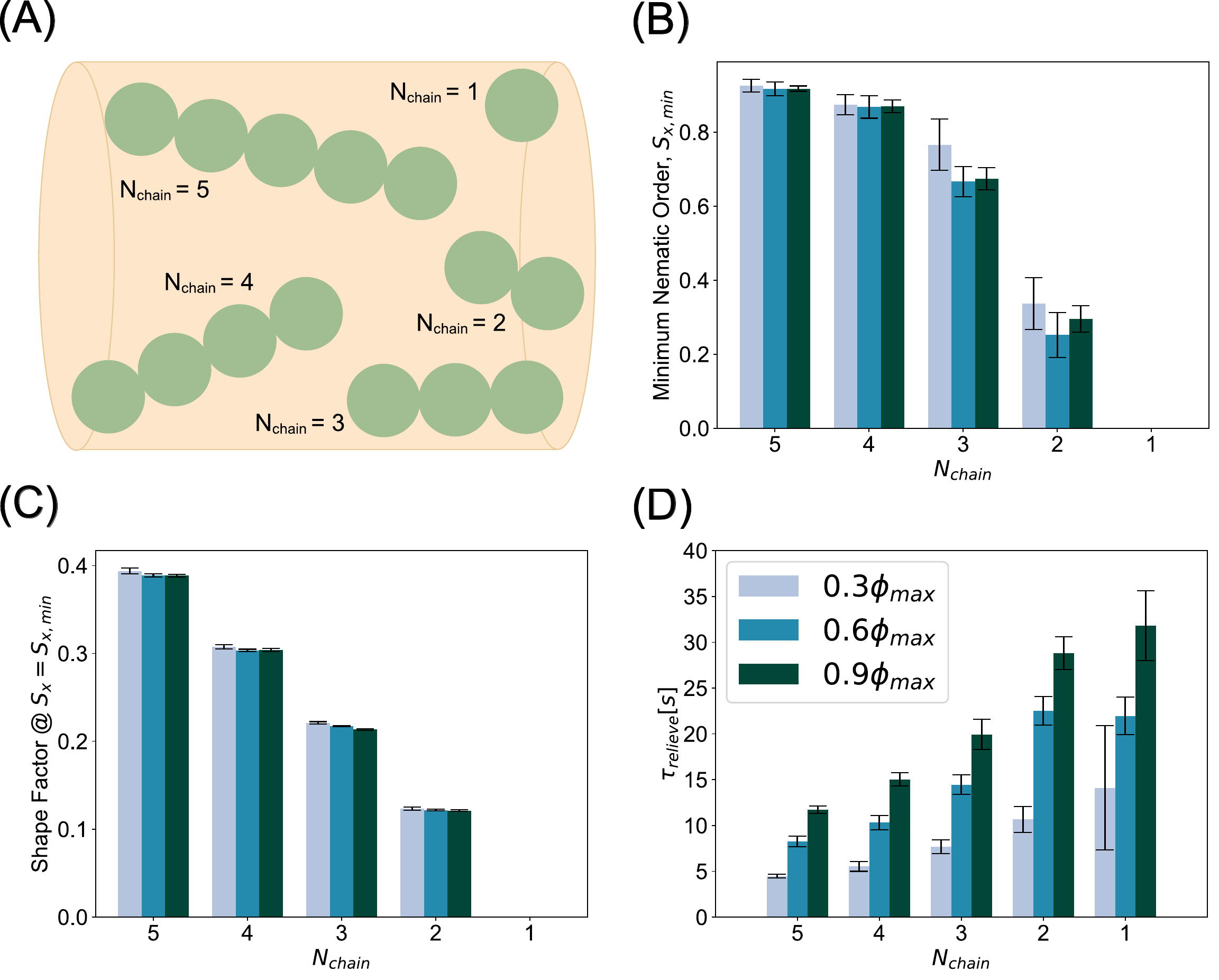}
    \caption{\textbf{Compact mitochondrial morphologies accentuate jamming severity.} (A) Schematic of mitochondrial morphology parameter, $N_\text{chain}$, the number of beads comprising bead spring chain. 
    (B) Minimum nematic order indicates shorter mitochondria become orientationally scrambled. 
    Conversely, longer mitochondria largely retain global orientational order even through jamming. (C) Shape factor at the time of minimum nematic order is largely insensitive to jamming across densities. (D) Across all densities, longer, more anisotropic mitochondria expedite transport recovery while shorter, isotropic morphologies prolong the jammed state. Bar heights and error bars in (B-D) represent the mean and standard deviation, respectively, of ten simulations per parameter.}
    \label{fig:ChainLength}
    \end{figure*}

We next asked whether mitochondrial morphology also contributes to transport disruption. 
Experimental observations suggest a link between mitochondrial morphology, mitochondrial density, and axonal morphology \cite{nikic2011reversible,mendelsohn2022morphological}. 
In addition, mitochondrial morphology is known to influence its functional properties including ATP synthesis rate \cite{garcia2023mitochondrial,venkatraman2023cristae}.
We varied the mitochondrial morphology along a continuum from granular spherical mitochondria $\left(N_\text{chain} = 1 \right)$ to elongated chain mitochondria $\left(N_\text{chain} = 5 \right)$ (\Cref{fig:ChainLength}A), where, for each value of $N_\text{chain}$, all mitochondria in that simulation were of that same length. 
We computed the value of the minimum nematic order parameter for mitochondria with bending stiffness $k_\text{b,mito} = 10^{-19} \ \text{Nm}^{2}$, while varying length of the mitochondria. 
We found that the global orientation of mitochondria of aspect ratio $\leq 3$ was significantly diminished by collisions, while those of aspect ratio $> 3$ retained global order throughout the collision (\Cref{fig:ChainLength}B). 
Next, we computed the shape factor at the time when the minimum nematic order occurs. 
We found that shape factor scales with chain length and is largely insensitive to density (\Cref{fig:ChainLength}C, \Cref{fig:Chain_SI}). 
Our previous analyses in \Cref{fig:fig4} suggested that conditions of mitochondria with the most resilient nematic order and shape factor would result in shortest relief times during traffic jams in axons. 
Consistent with this prediction, we found that across all densities tested, jamming was relieved most quickly by the highest aspect ratio chains. 
We found that mitochondria with high aspect ratio relieve jamming three to four times faster than their spherical counterparts. (\Cref{fig:ChainLength}D). 
Thus, our model predicts that mitochondrial accumulations in axons are enriched when mitochondria have isotropic morphologies with low aspect ratios (Movie 4). 

\subsection*{Mitochondrial life cycle affects jamming in axons}

Neuronal mitochondrial morphology is dynamic and is spatiotemporally controlled by fission and fusion \cite{westrate2014mitochondrial}. 
During mitochondrial fission, molecular machinery including proteins like Drp1,  bind to the outer mitochondrial membrane and serve as the scission site for a single mitochondrion to split into two.
In contrast, two mitochondria fuse into a single mitochondrion when molecules such as  mitofusin, MFN1/2, proteins fuse outer mitochondrial membranes and optic atrophy 1 (Opa1) proteins fuse inner membranes \cite{van2013mechanisms}.
The fission and fusion machinery establishes a life cycle for mitochondrial morphology \cite{westermann2010mitochondrial} and is regulated by upstream signaling \cite{khalilimeybodi2025systems}.
We next considered how the dynamics of fission and fusion can alter the jamming of mitochondria in axons. 
Based on our results thus far, we hypothesized that increased fission would promote prolonged jamming, while fusion can relieve jamming in axons and recover transport. 

\begin{figure*}[b!]
    \centering
    \includegraphics[width=\linewidth]
    {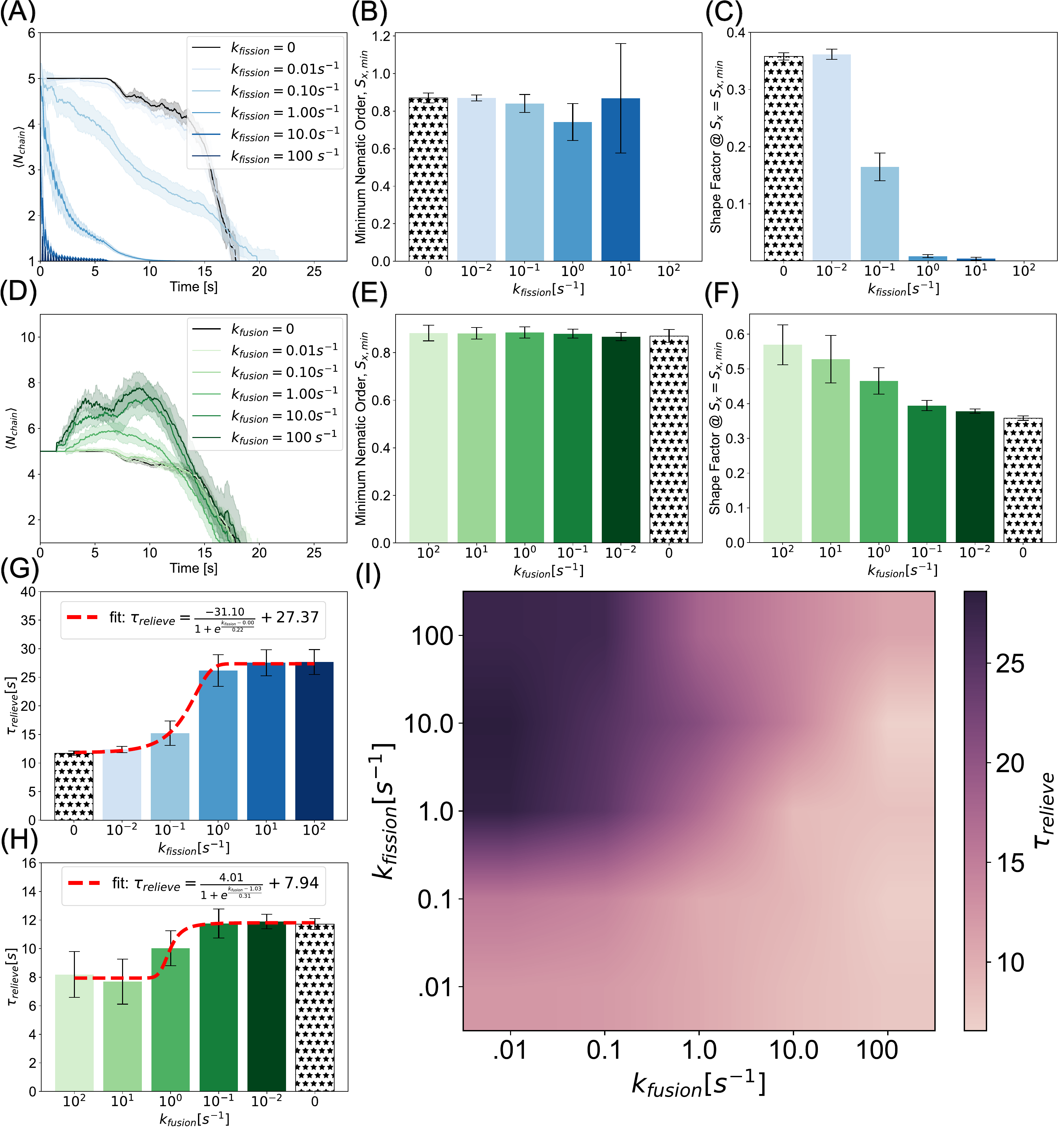}
    \caption{(Caption next page.)}
    \label{fig:FissionFigv2}
    \end{figure*}

We first implemented fission and fusion separately. 
The details of the implementation are given in Model Development.
We varied the fission and fusion rates as noted in 
\Cref{tab:figvar}.
The median of these values was motivated by \cite{arkfeld2025whole}, and we varied two orders of magnitude faster and slower to study the effect on jamming. 
To verify that our algorithm was implementing fission as expected, we calculated the ensemble averaged chain lengths of mitochondria as a function of time, where mitochondria were generated as $N_\text{chain} = 5$, and confirmed that fission gives rise to granular mitochondria (\Cref{fig:FissionFigv2}A). 
At low fission rates, $k_\text{fission} = 0.01 \ \text{s}^{-1} \ \text{and} \ k_\text{fission} = 0.1 \ \text{s}^{-1}$, orientational and anisotropic order were similar to the static case in \Cref{fig:ChainLength}B where $N_\text{chain} = 5$. 
At intermediate fission rates, $k_\text{fission} = 1 \ \text{s}^{-1} \ \text{and} \ k_\text{fission} = 10 \ \text{s}^{-1}$, the variance of the nematic order increased and, correspondingly, the shape factor decreased. 
This was due to the heterogeneous distribution of chain lengths produced by frequent fission events. 
At high fission rates, both the nematic order and shape factor tend toward zero as the population of mitochondria tends towards spherical morphologies (\Cref{fig:FissionFigv2}B,C, Movie 5). 

\addtocounter{figure}{-1}
\begin{figure*}[t!]    
    \caption{\textbf{A balance between fission and fusion controls jamming probability.} (A) Average chain length indicates high fission rates rapidly give rise to fragmented single bead mitochondria. (B) Nematic order remains close to one for slow fission. Fission gives rise to high variance regimes at sufficient rate constants ($k_\text{fission} > 1 \ \text{s}^{-1}$), when the population is predominantly $N_\text{chain} = 1$ and $N_\text{chain} = 2$ mitochondria that saturates to zero in the rapid fission limit. (C) Consistent with chain length analysis, the shape factor decreases as faster fission produces more granular morphologies. (D) Average chain length time courses show fusion produces more elongated chains than the initial condition morphology of $N_\text{chain} = 5$. (E) Minimum nematic order is largely insensitive to fusion rate as chains are long enough to expedite jamming relief even in the static case. (F) Shape factor increases, as expected, with fusion rates as longer chains are dynamically formed. (G,H) Relief times in fission and fusion rates can be described by sigmoidal functions saturating in exacerbated jamming at high and low values, respectively. (I) Consistent with the emergent morphological properties, phase diagram of fission rate versus fusion rate reveals that while fission serves as a mechanism to promulgate jamming via fragmentation, fusion can alleviate motility dysregulation. Curves and bar heights represent the mean of ten simulations per condition. Shaded regions and error bars represent the standard deviation of the same data.}
    \end{figure*}

We also implemented mitochondrial fusion as described in Model Development.
Our simulations showed that average chain lengths scale with fusion rate (\Cref{fig:FissionFigv2}D). Note that this distribution peaks and then decays as mitochondrial beads leave the simulated axon - shortening the chain of which it is leaving. 
We found that the minimum nematic order was insensitive to fusion rate.
This suggests that above a critical threshold mitochondrial length, the nematic order is not affected by collisions (\Cref{fig:FissionFigv2}E). Consistent with this observation, the shape factor increases with fusion rate (\Cref{fig:FissionFigv2}F), indicating that fusion promotes elongated conformations that remain aligned under crowding (Movie 6).

As was the case in our analyses of morphology dependence in the static conditions (\Cref{fig:fig4,fig:ChainLength}), we found that fragmented mitochondria that result from mitochondrial fission exhibit prolonged relief times, while long-chain tubular morphologies corresponding to fused mitochondria exhibit faster relief. 
As a result, we predict that fission-dominant conditions promote mitochondrial accumulation and persistent jamming. 
Consistent with our static results, we further predict that sites of elevated mitochondrial density are more likely in fission over-expressed mutants. 
Further, we predict that fusion promotes homeostatic transport, except in regions where geometric constraints become limiting, i.e., when axon diameter is smaller than mitochondrial length.
Interestingly, both of these trends are well-described by sigmoidal curves (\Cref{fig:FissionFigv2}G,H). 
Within the constraints of our model, this means that at the modeled axon length, fission rates below $k_\text{fission} = 1 \ \text{s}^{-1}$ are not fast enough to convert elongated mitochondria to granules to accentuate jamming. Similarly, fusion rates equal to and less than $k_\text{fusion} = 1 \ \text{s}^{-1}$ are insufficient to elongate chains prior to collisions. 

Finally, because fission and fusion take place simultaneously in axons, we simulated combined fission and fusion in our simulation framework to to construct a phase diagram of relief times
(\Cref{fig:FissionFigv2}I).
Our initial condition was the same as in the fission-only and fusion-only cases.
We found that increasing fission rates prolonged jamming, while increasing fusion rates alleviates it.
Furthermore, we found that a balance of fission and fusion can lead to shorter relief times. This suggests that as long as fission and fusion rates are comparable or biased slightly toward fusion, prolonged jams will not be likely. 
Indeed, fission and fusion rates have been observed to be identical in cortical
neurons and cerebellar granule neurons \cite{cagalinec2013principles}. 
An imbalance in mitochondrial fission and fusion rates is associated with impaired transport, as observed in 
mitofusin1 and mitofusin2 knockdown mutants of mouse embryonic fibroblasts \cite{chen2003mitofusins} and the failure of mitochondria in mitofusin knockdown mutants to distribute beyond the proximal somal region in neurons \cite{chen2007mitochondrial}. 

\subsection*{Mitochondrial jamming can lead to axonal swelling}

We next investigated the mechanical interactions of the axonal membrane-MPS composite and mitochondrial accumulations resulting from jamming.
Axons are not static structures \cite{costa2018regulation}. 
They are a mechanically complex structure \cite{bernal2007mechanical} comprised of a lipid bilayer decorated with transmembrane proteins including those which tether it to the cytoskeletal components in the cytoplasm \cite{xu2013actin}. This composite material is subject to both external and internal stresses that can result in axonal deformation. 
To simulate this deformation, we implemented the mechanics of the axon membrane-MPS composite (\Cref{fig:Schematic}D). 
In this implementation, we allowed the membrane nodes to be displaced according to the equations of motion described in module 3 of Model Development. 
\begin{figure*}[ht!]
    \centering
    \includegraphics[width=\linewidth]
    {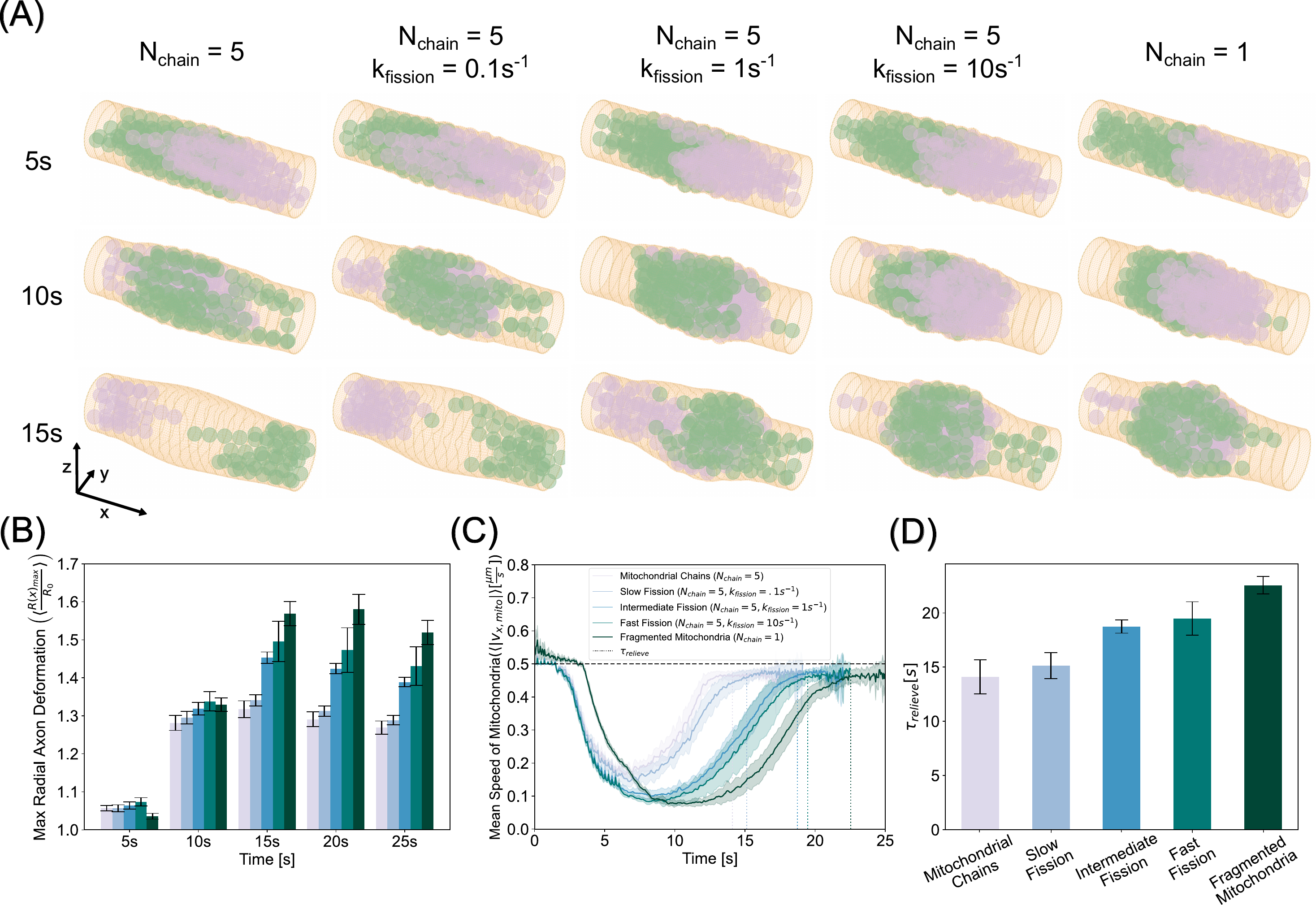}
    \caption{\textbf{Collision induced transport disruption stresses and deforms the axonal membrane.} (A) Simulation snapshots of initial anterograde-retrograde collisions (5s - top row), most highly jammed state (10s - middle row), and post collision (15s - bottom row) reveal that intermitochondrial collisions mutually hinder each others motion, thereby nucleating a site of accumulation that grows in time eventually exerting prominent forces on the membrane resulting in deformation. The left most column corresponds to the elongated mitochondria, the right column corresponds to granular mitochondria, and the three middle rows from left to right represent higher fission rates. (B) Maximum radial membrane deformations show a peak in time, with mitochondrial chains inducing ~30\% dilation and fragmented mitochondria producing up to nearly 60\% dilation. (C) Ensemble speeds of mitochondria show fragmented morphologies exhibit both a larger amplitude and period of stalling in deformed axons than their polymeric counterparts. (D) Relief times confirm granular morphologies are most susceptible to prolonged jamming residence times.}
    \label{fig:Deformations}
    \end{figure*}
We conducted simulations with stiff $\left( k_\text{b,mito} = 10^{-19} \ \text{Nm}^{2} \right)$ mitochondria of chain length $N_\text{chain} = 5$ that could not fuse but could undergo fission at rates $k_\text{fission} = 0 \text{\ (Movie 7)}, \ 0.1, \ 1, \ \text{and} \ 10 \ \text{s}^{-1}$ and granular mitochondria $\left( N_\text{chain} = 1 \right)$ (Movie 8). 
\Cref{fig:Deformations}A shows simulated axonal segments (membrane and spectrin - tan, actin rings - dark tan) with anterograde (green) and retrograde (plum) mitochondria at initial collision (top), peak jamming (middle), and post jamming (bottom). 

When mitochondria collide in the center of the axonal segment, the steric forces that govern mitochondrial interactions with the axonal membrane result in a local deformation in the membrane. As more mitochondria gather at the collision site, this accumulation exerts increasing radially divergent stresses on the membrane. These stresses stretch the membrane and increase the cross-sectional area. 
We found that elongated mitochondria deform the membrane less than their granular counterpart (\Cref{fig:Deformations}B). 
We then computed the time series of ensemble speeds and found, consistent with the fixed axon case, that elongated mitochondria recover their velocity faster than spherical mitochondria (\Cref{fig:Deformations}C). 
Indeed, we found that relief times mirror axonal deformation trends, where elongated mitochondria both induced less membrane stress and escaped jamming faster (\Cref{fig:Deformations}D). 
In contrast, spherical mitochondria induced larger axonal swellings due to their higher propensity to form persistent traffic jams and exert sustained forces on the membrane. 
Thus, we found a direct relationship between conditions that promote mitochondrial jamming and axonal swelling. 

\section{Discussion}

In this work, we have developed a minimal, biophysical model of motor-driven mitochondrial transport in axonal segments. 
Simulations from our model showed that mitochondrial collisions can nucleate jamming sites at locations where organelle transport is stifled. 
These traffic jams emerge from the balance of forces between self-propulsion and steric interactions, and their severity is modulated by mitochondrial morphology, mechanics, and density.
We further demonstrated that jamming severity can be exacerbated by isotropic, granular morphologies and increased density of mitochondria. 
We then established a predictive dependence of jamming severity on mitochondrial mechanics. 
More flexible mitochondria are prone to adopting globular morphologies that reinforce jamming.
In contrast, increased rigidity of these  organelles preserves elongation.
Thus, we show that elongated, rigid mitochondria behave like active nematics that can more easily pass by each other and reduce transport arrest.
Further, we established that the mitochondrial life cycle, particularly the balance between fission and fusion, strongly regulates collective transport.
When fusion and fission rates are comparable, jamming is minimized.
Fission-dominant conditions promote transport disruption and fusion-dominant conditions facilitate recovery of motility.
We found relief times as functions of fission and fusion were both well fit by sigmoidal curves. Sigmoidal functions are ubiquitous in biological systems. Often referred to as ``dose response curves'', sigmoidal functions occur across  scales from the Hill equation describing ligand binding dynamics \cite{gesztelyi2012hill} to the logistic function describing population dynamics. 
This suggests a lifecycle dynamic rate centered around a high sensitivity regime wherein relative rates can be finely tuned for large effect to safeguard transport fidelity. 
Mitochondrial transport disruption can also result in axonal swelling because of interactions with the axonal membrane.
The magnitude of deformation is proportional to the severity of the jam, thus indicating a direct link between mitochondrial transport disruption and axonal swelling.

Several independent experimental observations support these predictions. 
Comparable fission and fusion rates have been reported in both cortical
neurons and cerebellar granule neurons \cite{cagalinec2013principles}. 
In contrast, increased fission or reduced fusion is associated with heterogeneous mitochondrial distributions and impaired transport as observed in Alzheimer's models  \cite{wang2009impaired}. Mitofusin1 and mitofusin2 knockouts of mouse embryonic fibroblasts \cite{chen2003mitofusins} show reduced motility. 
In Purkinje cells, mitofusin knockdown prevents mitochondria from escaping the proximal somal region into dendritic branches and alters the mitochondrial distribution and dynamics \cite{chen2007mitochondrial}.
It has also been observed, however, that Drp1 knockouts in Purkinje cells do not fully distribute themselves in the dendritic arbor \cite{fukumitsu2016mitochondrial}. 
While that is not a condition directly tested in our model, the biophysical principles we have discovered suggest that 
fission knockdowns can give rise to large mitochondrial morphologies whose size would prohibit them from occupying small dendritic branches. 
Axonal dilations, termed as ``focal axonal swellings", have been observed across a broad swath of experimental systems that analyze neurological conditions in both myelinated and unmyelinated cells, with little mechanistic insight into the origin of the deformations \cite{nikic2011reversible,tang2012partial}. 
Mitochondrial accumulation has been identified in the swelling sites of mouse sciatic nerves and cat retinas \cite{greenberg1990irregular}, mouse optic nerves \cite{wang2011traumatic}, and TgCRND8 amyloid precursor protein transgenic mice axons \cite{adalbert2009severely}.
Consistent with our model, mouse models of multiple sclerosis have shown mitochondrial aspect ratio distributions centered closer to one in axonal swellings than their cylindrical axon counterparts \cite{nikic2011reversible}. 
This supports our prediction that granular mitochondrial morphologies are capable of inducing increased deformations of the axonal membrane.


It has been known for over two decades, now, that axonal damage is a potent predictor of a host of neurological conditions \cite{medana2003axonal}. 
Axonal swelling density is linked to progression rates of patients with painful neuropathy \cite{lauria2003axonal}. 
It has also been shown that induced axonal swellings are followed by mitochondrial accumulation due to broken microtubules at the site of acute axon dilation \cite{tang2010mechanical}. 
Here, we show a different mechanism of axon deformation that can result from mitochondrial accumulation, suggesting the possibility of the existence of a feedback loop between mitochondrial accumulation and axonal swelling. 
Thus, a specific experimental prediction from our work would be to induce a high density of mitochondria in a homeostatic axon to test whether mitochondria-mitochondria and mitochondria-membrane interactions during transport can, indeed, initiate membrane deformation.


One important prediction from the model is that mitochondrial bending rigidity can affect jamming. 
Our simulations reveal that rigid mitochondria will jam less than their flexible counterparts. 
While direct measurements of mitochondrial membrane rigidity are not available, it is well-established that mitochondrial lipid composition can alter membrane properties \cite{venkatraman2023cristae,Lee2025-mm}.
It has also been shown that mitochondria membrane deformability is pH-dependent \cite{wang2008membrane}. 
Therefore, experiments that can perturb mitochondrial membrane mechanics (lipid composition or osmotic pressure changes) might be able to test our model predictions. 
Future studies might also look at how axons which have pearls-on-a-string morphology \cite{griswold2025membrane} might influence jamming of mitochondria. 

The primary objective of this study was to investigate dystrophic behavior of mitochondrial transport arising from collective mitochondrial interactions.
We modeled the action of kinesin and dynein motors on mitochondria as a highly coarse-grained self-propulsion. 
In reality, the number of coupled interactions between motors, cytoskeleton, and mitochondria that result in neuronal transport is much more complicated \cite{pekkurnaz2022mitochondrial,kruppa2021motor}. The axoplasm is an extremely crowded environment. Mitochondrial mechanical plasticity \cite{fischer2018morphology} and direct and indirect adaptor protein interactions \cite{henrichs2020mitochondria} facilitate transport, maintaining function even through cytoskeletal obstructions \cite{balint2013correlative}. 
Future efforts will focus on explicit considerations of motor proteins and microtubule interactions with the mitochondria. 
Integrating metabolic feedback into the present framework would enable direct testing of how transport, morphology, and energy production co-regulate mitochondrial positioning.
Thus, our biophysical model links organelle organization to axonal pathology and not only provides testable predictions for how imbalances in fission and fusion drive structural and functional failure in neurons but also provides opportunities for building increasingly detailed mechanistic framework.

\newpage
\printbibliography

\section{Acknowledgments}
This work was funded by the National Institutes of Health (NIH) under grants 5R01MH139350-02 (to P.R.), R35GM128823 and R01NS136048 (to G.P.), and 5T32GM133351 (to A.A.A.). We acknowledge the Triton Shared Computing Cluster (TSCC) at the San Diego Supercomputing Center (SDSC) on which the majority of the simulations were run. 
Portions of the simulation and analysis code were reviewed, and debugged by members of the Rangamani lab and Google Gemini 3 Flash.

\section{Competing Interests}
P.R. is a consultant for Simula Research Laboratories in Oslo, Norway and receives income. The terms of this
arrangement have been reviewed and approved by the University of California, San Diego in accordance with its
conflict-of-interest policies. The remaining authors declare no competing interests.




\section{Supplementary Material} 
\setcounter{figure}{0}
\setcounter{table}{0}
\renewcommand{\thetable}{S\arabic{table}}
\renewcommand{\thefigure}{S\arabic{figure}}

\begin{figure*}[h]
    \centering
    \includegraphics[width=\linewidth]{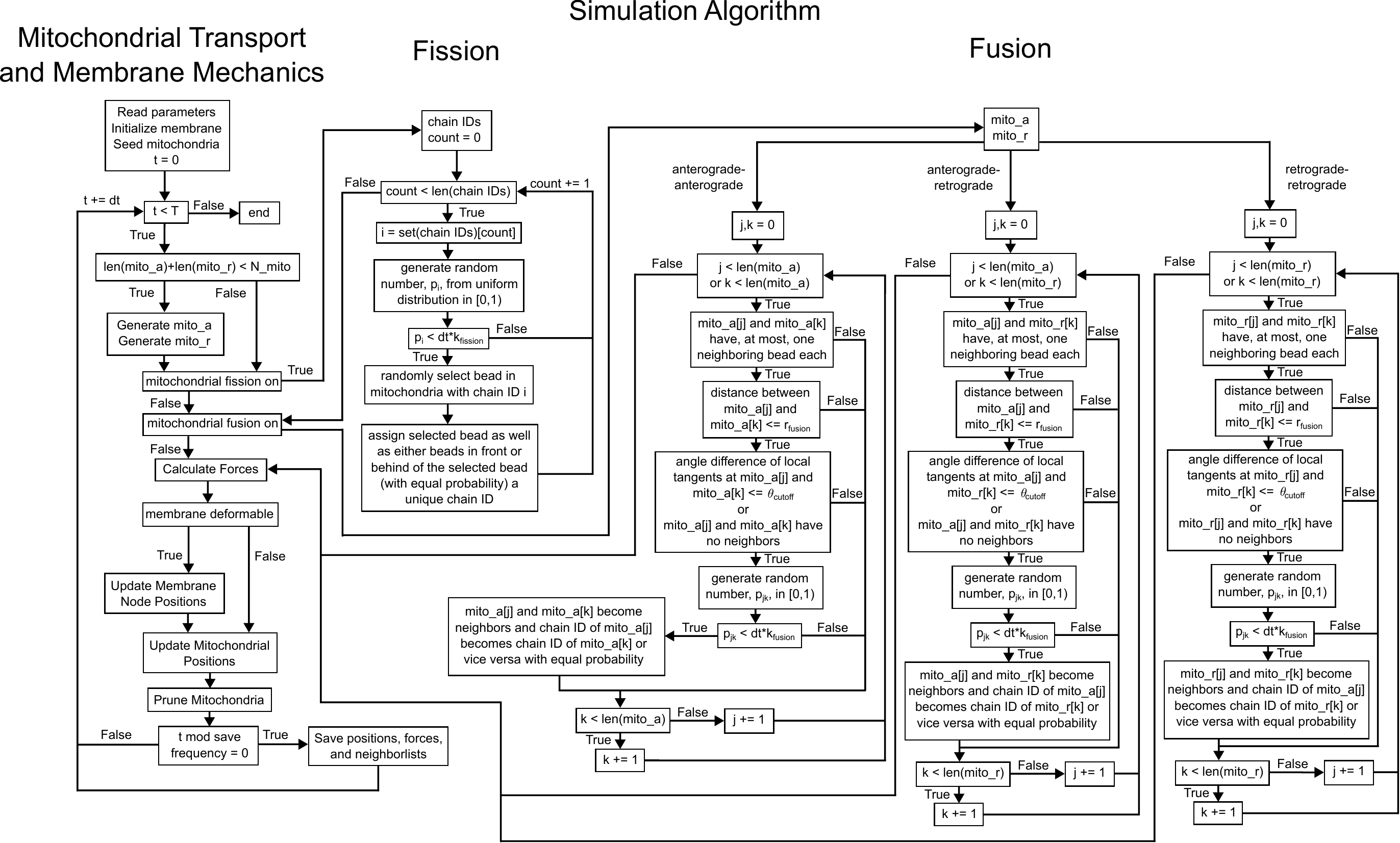}
    \caption{Graphical Algorithm of Mitochondrial transport in axon - including mitochondrial transport and membrane mechanics (left), mitochondrial fission (middle), and mitochondrial fusion (right).}
    \label{fig:Algorithm}
\end{figure*}

\begin{table*}[h]
    \centering
    \caption{Figure Parameter Variations}
    \begin{tabular}{ | m{2cm} | m{8cm}| m{5cm} |} 
        \hline
        Figure & Varied Parameter(s) & Values \\
        \hline
        1 & - & - \\
        \hline
        2 & Density $\left( \phi \right) [1]$ & 0.3$\phi_\text{max}$, 0.6$\phi_\text{max}$, 0.9$\phi_\text{max}$ \\
        \hline
        3 & Mitochondrial bending rigidity $\left(k_\text{b,mito}\right) [\text{Nm}^{2}]$ \newline Density $\left( \phi \right) [1]$ & $10^{-21},10^{-20},10^{-19}$ \newline 0.3$\phi_\text{max}$, 0.6$\phi_\text{max}$, 0.9$\phi_\text{max}$ \\
        \hline
        4 & Number of beads per mitochondria $\left(N_\text{chain}\right)[\#]$ \newline Density $\left( \phi \right) [1]$ & 1, 2, 3, 4, 5 \newline 0.3$\phi_\text{max}$, 0.6$\phi_\text{max}$, 0.9$\phi_\text{max}$ \\
        \hline
        5 & Fusion rate $\left(k_\text{fusion}\right) [\text{s}^{-1}]$ \newline Fission rate $\left(k_\text{fission}\right) [\text{s}^{-1}]$ & 0, 0.01, 0.1, 1.0, 10, 100 \newline 0, 0.01, 0.1, 1.0, 10, 100 \\
        \hline
        6 & Fission rate $\left(k_\text{fission}\right) [\text{s}^{-1}]$ \newline Number of beads per mitochondria $\left(N_\text{chain}\right)[\#]$ & 0, 0.1, 1, 10 \newline 1, 5 \\
        \hline
    \end{tabular}
    \label{tab:figvar}
\end{table*}

\noindent Movie 1: Low Density ($\phi = 0.3\phi_\text{max}$), low stiffness ($k_\text{b,mito}=10^{-21} \ \text{Nm}^{2}$), $N_\text{chain} = 5$ characteristic simulation. Jamming is minimal and anterograde and retrograde populations can pass by one another easily.

\noindent Movie 2: High Density ($\phi = 0.9\phi_\text{max}$), low stiffness ($k_\text{b,mito}=10^{-21} \ \text{Nm}^{2}$), $N_\text{chain} = 5$ characteristic simulation. Jamming is prominent and anterograde and retrograde populations require extended time to pass by one another. 

\noindent Movie 3: High Density ($\phi = 0.9\phi_\text{max}$), high stiffness ($k_\text{b,mito}=10^{-19} \ \text{Nm}^{2}$), $N_\text{chain} = 5$ characteristic simulation. Jamming is prevalent but the persistent anistropy of the mitochondrial chains aid in alleviating the jam.

\noindent Movie 4: High Density ($\phi = 0.9\phi_\text{max}$) $N_\text{chain} = 1$ characteristic simulation. Jamming is prominent and the granularity of the mitochondria reinforce the jamming severity.

\noindent Movie 5: High Density ($\phi = 0.9\phi_\text{max}$), high fission rate ($k_\text{fission}=100 \ \text{s}^{-1}$), $N_\text{chain} = 5$ characteristic simulation. Jamming is prominent and plentiful fission events resulting in granular mitochondria reinforce the jamming severity. 

\noindent Movie 6: High Density ($\phi = 0.9\phi_\text{max}$), high fusion rate ($k_\text{fusion}=100 \ \text{s}^{-1}$), $N_\text{chain} = 5$ characteristic simulation. Jamming is minimized as plentiful fusion events give rise to highly anisotropic tubular mitochondria that can easily pass by one another.

\noindent Movie 7: High Density ($\phi = 0.9\phi_\text{max}$), $N_\text{chain} = 5$, with membrane deformations characteristic simulation. Elongated mitochondrial morphologies exert less stresses on the membrane than fragmented spherical mitochondria.

\noindent Movie 8: High Density ($\phi = 0.9\phi_\text{max}$), $N_\text{chain} = 1$, with membrane deformations characteristic simulation. Membrane dilation at the collision site locally increase the cross sectional area, providing more paths for mitochondria to escape the jamming.


\begin{figure*}
    \centering
    \includegraphics[width=\linewidth]
    {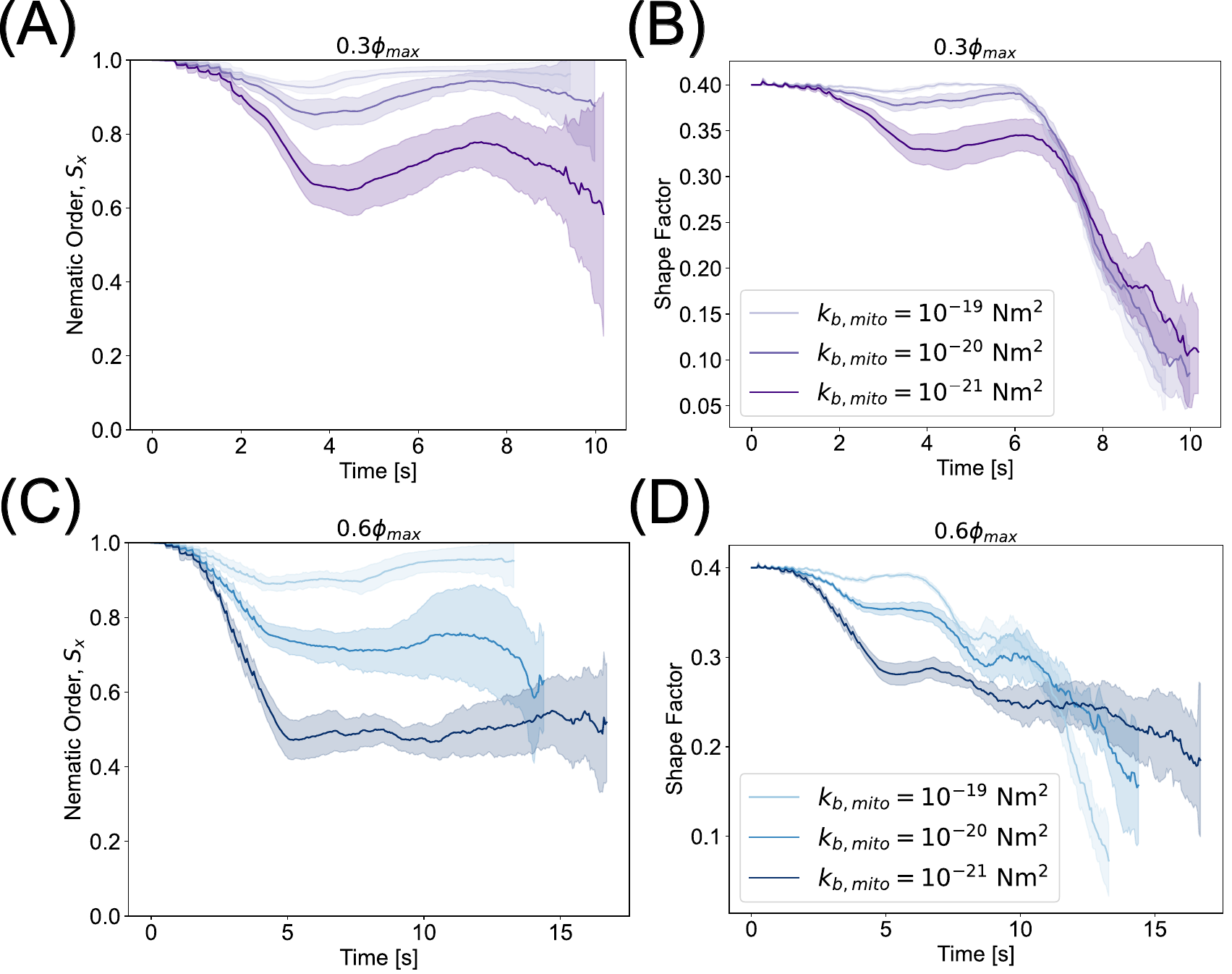}
    \caption{\textbf{Nematic and shape factor analysis of intermediate and sparse densities for varying bending stiffness.} (A,B) Nematic order and shape factor of lowest densities, $\phi = 0.3\phi_\text{max}$, shows $10^{-19} \ \text{Nm}^{2}$ and $10^{-20} \ \text{Nm}^{2}$ bending rigidities recover initial morphologies, while $10^{-21} \ \text{Nm}^{2}$ only partially recovers. (C-D) Intermediate densities exhibit very similar behavior to their high density counterparts described in the main text. Curves and shaded regions represent the mean and standard deviation, respectively, of ten simulations per parameter set.}
    \label{fig:Fig4_SI}
\end{figure*}

\begin{figure*}
    \centering
    \includegraphics[width=\linewidth]{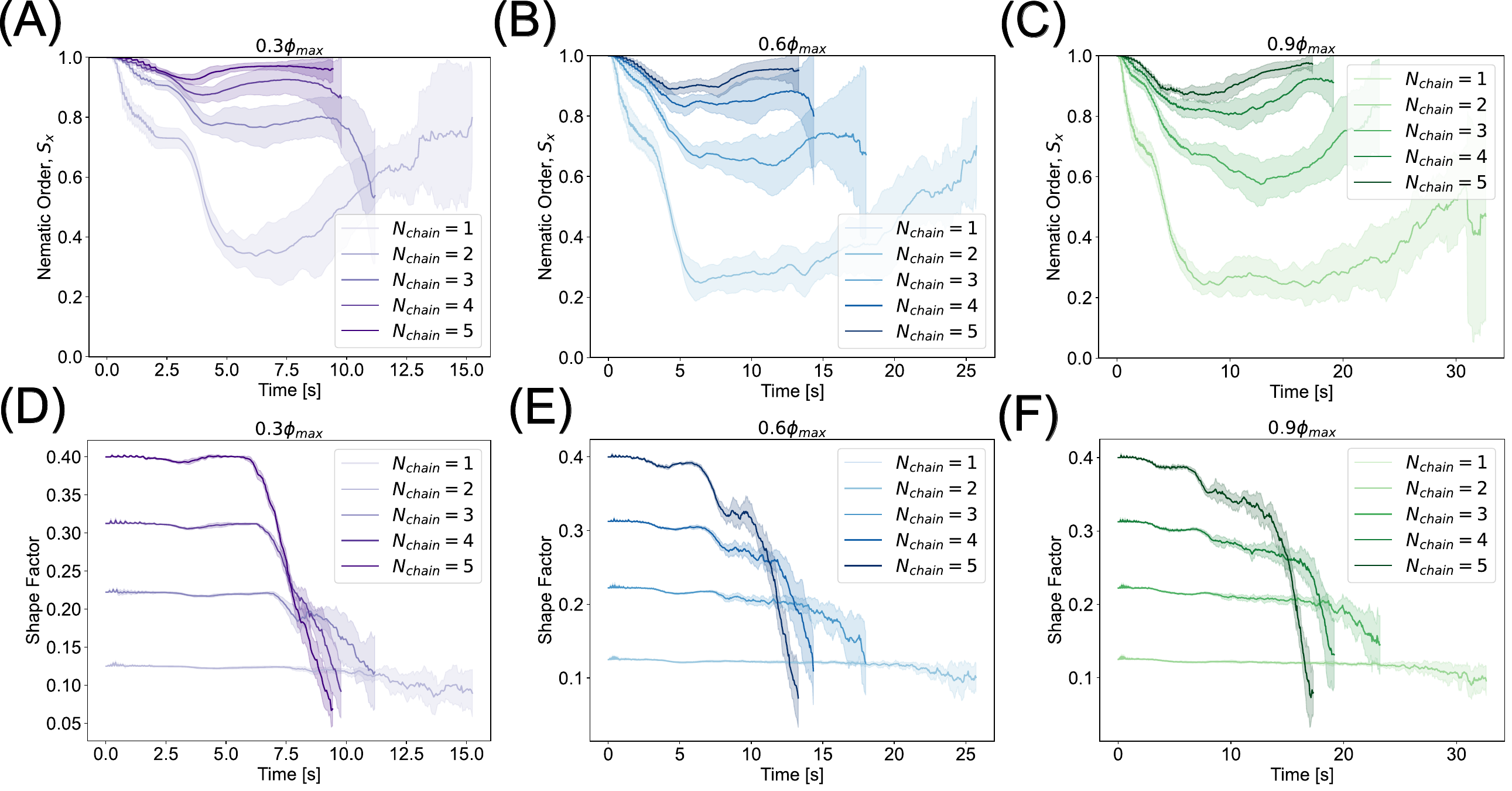}
    \caption{\textbf{Nematic and shape factor analysis of intermediate and sparse densities for varying mitochondrial aspect ratio.} (A-C) Nematic order time courses of increasing density for morphologies from granular $\left( N_\text{chain} = 1 \right)$ to elongated $\left( N_\text{chain} = 5 \right)$. Elongated mitochondria retain nematic order and recover unity while shorter morphologies show a reduced orientational order that does not recover. (D-F) Corresponding shape factors. Elongated mitochondria have a higher shape factor than shorter counterparts. Shape factor decays near the end of the trajectory as beads are exiting the simulation axon. Curves and shaded regions represent the mean and standard deviation, respectively, of ten simulations per parameter set.}
    \label{fig:Chain_SI}
\end{figure*}



\end{document}